\documentclass[journal=ancac3,manuscript=article]{achemso}

\usepackage[version=3]{mhchem} % Formula subscripts using \ce{}
\usepackage{graphicx} % Required to include images
\usepackage{caption}
\usepackage{subcaption}
\usepackage{mathtools}
\usepackage{amsfonts}
\author{Lara Rom\'an Castellanos}
\affiliation{Department of Physics, Imperial College London, London SW7 2AZ}
\author{Ortwin Hess}\email{ortwin.hess@tcd.ie}
\affiliation{Department of Physics, Imperial College London, London SW7 2AZ}
\alsoaffiliation{The Thomas Young Centre for Theory and Simulation of Materials}
\alsoaffiliation{School of Physics and CRANN Institute, Trinity College Dublin, Dublin 2, Ireland}
\author{Johannes Lischner}\email{jlischner@imperial.ac.uk}
\affiliation{Department of Physics and Materials, Imperial College London, London SW7 2AZ}
\alsoaffiliation{The Thomas Young Centre for Theory and Simulation of Materials}

\title{Dielectric engineering of hot carrier generation by quantized plasmons in embedded silver nanoparticles}

\begin{document}


\begin{abstract}
Understanding and controlling properties of plasmon-induced hot  carriers  is  a  key step towards next-generation photovoltaic and photocatalytic devices. Here, we uncover a route to engineering hot-carrier generation rates of silver nanoparticles by designed embedding in dielectric host materials. Extending our recently established quantum-mechanical approach to describe the decay of quantized plasmons into hot carriers we capture both external screening by the nanoparticle environment and internal screening by silver d-electrons through an effective electron-electron interaction. We find that hot-carrier generation can be maximized by engineering the dielectric host material such that the energy of the localized surface plasmon coincides with the highest value of the nanoparticle joint density of states. This allows us to uncover a path to control the energy of the carriers and the amount produced, for example a large number of relatively low-energy carriers are obtained by embedding in strongly screening environments.
\end{abstract}


\section{Introduction}

Understanding and controlling light-matter interactions at the nanoscale is important for increasing the efficiency of photovoltaic and photocatalytic devices~\cite{Saavedra2016,Molinari2017,Hartland2017}. In this context, localized surface plasmons (LSP) in metallic nanoparticles provide a unique platform because the LSP decay generates energetic or "hot" carriers that can be harnessed to induce chemical reactions~\cite{Li2016} or overcome interfacial barriers~\cite{Clavero2014}. Other applications of hot carriers include biochemical sensing~\cite{Anker2008}, spectroscopy~\cite{Halas2011}, nanophotonic lasers~\cite{Wang2018} and quantum information devices~\cite{Kolwas2019}. These applications usually require a generation of a large number of energetic carriers, but in many current devices the hot-carrier generation rates are relatively small~\cite{Liu2019}. 

To improve the performance of hot-carrier devices, significant efforts have been made to understand the dependence of hot-carrier properties on the nanoparticle material and its geometry~\cite{Besteiro2017,Naldoni2017,Brown2015}. However, in real devices the nanoparticles are often embedded in insulating host materials or placed on top of their surfaces~\cite{Raza2015,Mittal2015,Crut2014,Codrington2017} and therefore the influence of the nanoparticle environment on hot carriers must also be considered. It is well known that the dielectric properties of the nanoparticle environment modify the LSP frequency~\cite{Kreibig1987}, influence interfacial transport barriers and protect the nanoparticle from oxidation~\cite{Zong2005}, but not much is known about their effect on hot-carrier generation rates. 

Microscopically, the dielectric environment modifies the effective interaction between conduction electrons in the nanoparticle. To approximate this screened interaction, several groups have modelled the environment as a linear polarizable medium, solved the corresponding Maxwell equations for a point charge in this system and used the result to study the changes in the photoabsorption behaviour of the nanoparticle induced by the environment~\cite{Raza2015,Rubio1993,Campos2019}. By treating the nanoparticle itself as a polarizable medium, this approach can easily be extended to also capture the dielectric screening by bound charges in the material, such as d-band electrons in silver.

In this paper, we present a quantum-mechanical approach for calculating the effect of a dielectric environment on hot-carrier properties in embedded silver nanoparticles. In particular, we combine a recently developed method for describing hot-carrier generation by quantized plasmons~\cite{RomanCastellanos2019} with a screened interaction between conduction electrons that takes the dielectric response of the environment and also of the polarizable d-band electrons into account. We present results for four different host materials (silicon dioxide, titanium dioxide, silicon nitride and gallium phosphide) and compare them to results obtained for nanoparticles in air. We also study the dependence of hot-carrier properties on the size of the embedded nanoparticle. Our calculations reveal that hot-carrier rates in these systems depend sensitively on the dielectric properties of the environment. In particular, the environmental screening reduces the LSP energy and thereby changes the accessible LSP decay channels. Moreover, screening reduces the electron-plasmon coupling, but this effect can be compensated by the increase of the coupling due to the reduced LSP energy. These results pave the way towards a detailed understanding of hot-carrier generation in embedded nanoparticles and open up the possibility of dielectric engineering of hot-carrier properties.

\section{Methods}

We review here the solution of the Poisson equation for a sphere of radius $R$ with dielectric constant $\epsilon_d$ embedded in a material with dielectric constant $\epsilon_1$~\cite{Deng2009,Kirkwood1934, Serra1997}. As the charge density of a jellium nanoparticle can spill beyond the positive charge background representing the ions, three scenarios are considered: (i) the potential inside the nanoparticle generated by a charge inside the nanoparticle (denoted $V_{\text{in-in}}$), (ii) the potential outside the nanoparticle generated by a charge inside (denoted $V_{\text{in-out}}$) and (iii) the potential outside the nanoparticle generated by a charge outside the nanoparticle (denoted $V_{\text{out-out}}$). Note that the potential inside the nanoparticle generated by a charge outside is also described by $V_{\text{in-out}}$. We carry out linear response TDDFT calculations where the Coulomb interaction is described via these potentials which are given by
\begin{align}
\label{eq: VPol}
& V_{\text{in-in}}(\textbf{r},\textbf{r}')=
\frac{1}{\epsilon_{\rm d}|\textbf{r} - \textbf{r}'|} + \frac{1}{\epsilon_{d}} \sum^{\infty}_{l=0}  \frac{(\epsilon_{\rm d} - \epsilon_{1})\,(l+1)}{(l\epsilon_{d} + l\epsilon_{1} + \epsilon_{1})} \frac{(r r')^{l}}{R^{2l + 1}}  P_{l}(\cos\theta)   \; \; & r,r'  \leq R, \\ \nonumber \\ 
\label{eq: VPolinout}
& V_{\text{in-out}}(\textbf{r},\textbf{r}')= \frac{1}{\epsilon_{d}} \sum^{\infty}_{l=0} \frac{(2l+1)\epsilon_{\rm d}}{(l\epsilon_{\rm d} + l\epsilon_{1} + \epsilon_{1})} \frac{(r')^{l}}{r^{l+1}} P_{l}(\cos\theta)   \; \;  & r \ge R,r'\leq R, \\ \nonumber \\ 
\label{eq: VPoloutout}
& V_{\text{out-out}}(\textbf{r},\textbf{r}')= \frac{1}{\epsilon_{\rm 1}|\textbf{r} - \textbf{r}'|} + \frac{1}{\epsilon_{1}}   \sum^{\infty}_{l=0} \frac{l(\epsilon_{1} - \epsilon_{d})}{(l\epsilon_{d} + l\epsilon_{1} + \epsilon_{1})}\frac{R^{2l+1}}{r^{l+1}(r')^{l+1}}  P_{l}(\cos\theta) \; \; \; \;  & r,r' \ge R,
\end{align}
where $P_l$ denotes the Legendre polynomial of order $l$. In principle, the radius of the dielectric sphere is an adjustable parameter~\cite{Campos2019}. In our work, we set the radius of the dielectric sphere equal to the radius of the sphere of positive background charge of the jellium nanoparticles. It is also worth noting that the dielectric properties of real materials are frequency-dependent. Here, we neglect this frequency dependence (as this would drastically increase the cost of solving the Casida equation) and use the static electronic dielectric constants of the dielectric sphere and the environment in the effective interaction. The Coulomb integrals were computed using the LIBERI library \cite{Toyoda2010}. The TDDFT calculations were carefully converged with respect to the number of empty states. Note that our framework includes a quantized treatment of the plasmon, relevant to describe quantum effects present in small nanoparticles and/or when a low density of plasmons are excited as introduced in Ref.~\citenum{RomanCastellanos2019}.

%\textcolor{red}{and, as described by other authors in the literature, the LSP quantum annihilation generates a single electron-hole pair at a given time~\cite{Khurgin2020a}}.

\section{Results and discussion}
\subsection{Description of the model}

We have studied hot-carrier properties of embedded spherical silver nanoparticles consisting of between 68 and 254 atoms corresponding to diameters between 1.08 and 2.10~nm. To calculate hot-carrier distributions in these systems, we extended the approach developed in Ref.~\citenum{RomanCastellanos2019} for alkali metal nanoparticles. Following this approach, the decay of the localized surface plasmon (LSP) into a single electron-pair is considered~\cite{Khurgin2020a, Bernardi2015}. The corresponding generation rate $N(E)$ of hot electrons with energy $E$ created by the decay of a single LSP quantum is obtained from Fermi's golden rule according to
\begin{equation}
N(E) = \frac{2\pi}{\hbar} \sum_{vc} |g_{vc}|^{2} \delta(\epsilon_{c} - \epsilon_{v} - \hbar \omega_P) \delta(E-\epsilon_c),
\label{eq:Fermi}
\end{equation}
where $\epsilon_c$ and $\epsilon_v$ denote the quasiparticle energies of occupied and empty states 
and $\omega_P$ is the LSP frequency. To determine these quantities, we first carry out density-functional theory (DFT) calculations on jellium spheres with a Wigner-Seitz radius of $r_s=3.0$~Bohr (corresponding to the density of conduction electrons in the Ag sp-band) using the local density approximation~\cite{Kohn1965, Perdew1981}. Next, we use the $\Delta$-SCF approach to calculate the ionization potential of the nanoparticles~\cite{RomanCastellanos2019}. The quasiparticle energies are then obtained by shifting the Kohn-Sham (KS) energies by the difference between the ionization potential and the KS energy of the highest occupied orbital. Finally, the LSP frequency is obtained from time-dependent density-functional theory (TDDFT) calculations in the random-phase approximation.

In Eq.~\eqref{eq:Fermi}, the electron-plasmon coupling $g_{vc}$ is given by 
\begin{equation}
 g_{vc} = e^2\int d\mathbf{r} \int d\mathbf{r}' \phi_c(\mathbf{r})\phi_{v}(\mathbf{r})V(\mathbf{r},\mathbf{r'})\rho_{P}(\mathbf{r}'),
\label{eq:coupling}
\end{equation}
where $\phi_v(\mathbf{r})$ [$\phi_c(\mathbf{r})$] denotes the quasiparticle wavefunction of an occupied (empty) state, $\rho_P(\mathbf{r})$ is the LSP transition density and $V(\mathbf{r},\mathbf{r}')$ denotes the screened interaction between electrons. 

To describe the screening by the Ag d-electrons and by the nanoparticle environment, we calculate the potential created by a point charge in a sphere with dielectric constant $\epsilon_d$ surrounded by an environment with dielectric constant $\epsilon_1$. Solving the corresponding Poisson equation yields the effective interaction given in Eqs.~\eqref{eq: VPol}-\eqref{eq: VPoloutout}~\cite{Rubio1993,Campos2019}. Besides Eq.~\eqref{eq:coupling}, we also use this screened interaction to calculate the TDDFT interaction matrix elements and thereby capture the screening-induced changes to the LSP energy and transition density. For the internal screening by the d-band electrons, we use $\epsilon_d=3.3$~\cite{Romaniello2005} (in the Supplementary Information we verify that our results do not qualitatively depend on the precise value of this parameter). Note that the frequency-dependent screening of the conduction electrons is captured through the explicit solution of the Casida equation which yields the frequency-dependent susceptibility of the nanoparticle.

We expect that this approach for calculating hot-carrier properties in embedded silver nanoparticles gives accurate results as long as the LSP energy is smaller than the separation between the d-bands and the Fermi level which is approximately 4~eV \cite{Cazalilla2000}. If this condition is not fulfilled, the LSP decay can lead to the creation of hot holes in the d-bands and bandstructure methods beyond the jellium model are needed~\cite{RomanCastellanos2019}. In practice, we find that LSP energies in Ag nanoparticle are in fact smaller than 4~eV as long as internal screening by d-band electrons is taken into account.

Note that in the above equations $\hbar$ denotes the reduced Planck constant and $e$ is the electron charge. Also, we replace the two delta-functions in Eq.~\eqref{eq:Fermi} by Gaussians in our numerical calculations. The standard deviation of the first Gaussian is 0.12~eV reflecting the lifetime of the LSP~\cite{Sonnichsen2002} while the standard deviation of the second Gaussian is 0.05~eV reflecting the quasiparticle lifetime in Ag~\cite{Aeschlimann1996,Rossi2017b}.

Finally, it is worth pointing out that Eq.~\eqref{eq:Fermi} assumes ground-state occupancies, i.e. all states below the Fermi level are fully occupied and all states above the Fermi level are completely empty. Of course, different occupancies must be used if the nanoparticle contains an excited population of hot carriers which can result in the Pauli blocking of certain transitions. This can happen, for example, when the system is illuminated continuously by a strong light source or when electron-phonon coupling leads to a significant increase in the nanoparticle temperature~\cite{Dubi2019b}.

\subsection{Optical properties of silver nanoparticles}

\begin{figure}[h!]
\centering
\includegraphics[width=0.6\textwidth]{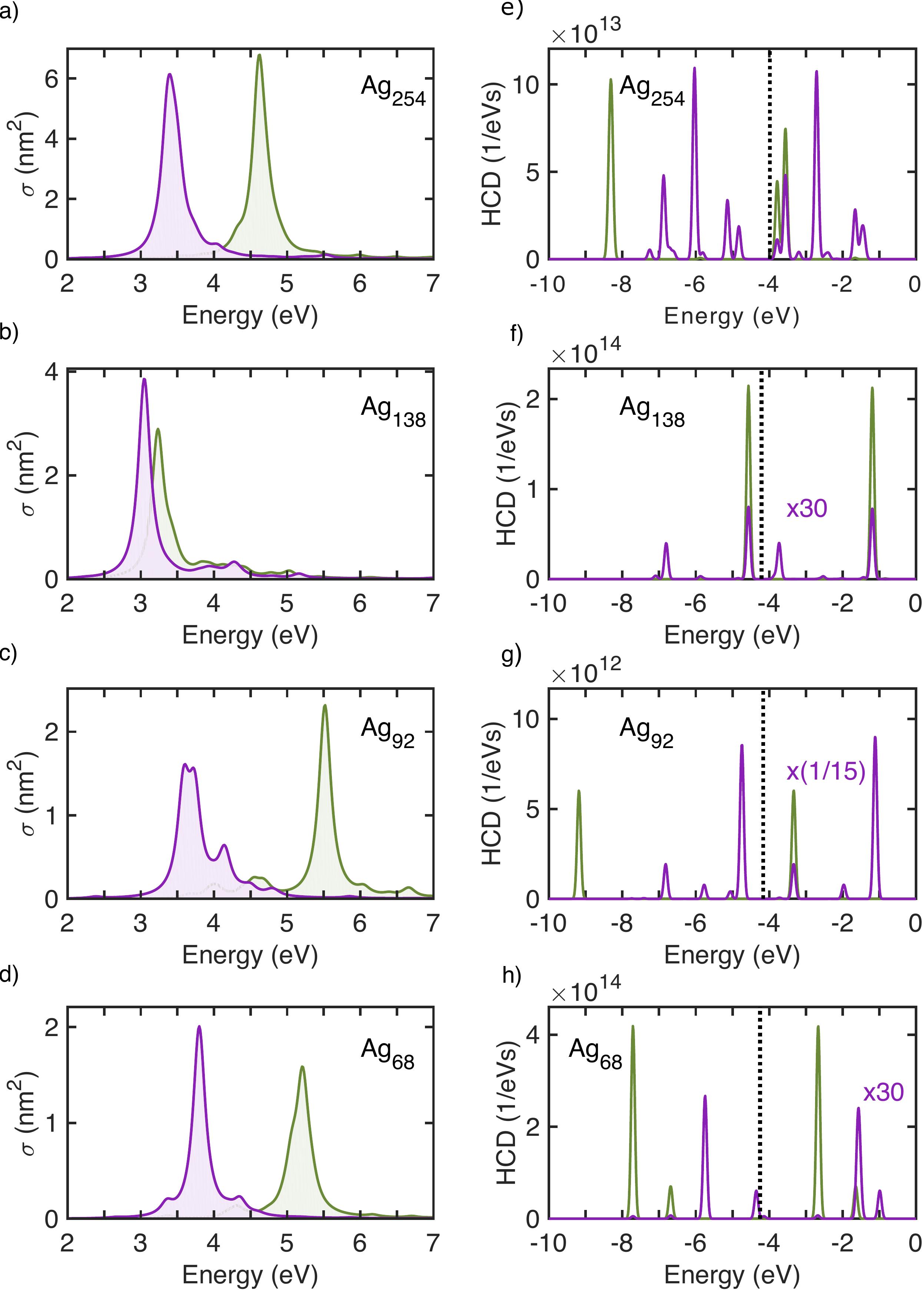}
\caption[Absorption spectra and hot carrier distributions for the silver nanoparticles with and without the effect of the d-electrons.]{Left: Absorption spectra of Ag nanoparticles in air with (purple curves) and without (green curves) d-electron screening as function of photon energy: (a) Ag$_{254}$, (b) Ag$_{138}$,(c) Ag$_{92}$ and (d) Ag$_{68}$. Right: Plasmon-induced hot carrier distributions (HCD) of Ag nanoparticles in air with and without d-electron screening as function of hot-carrier energy:(e) Ag$_{254}$, (f) Ag$_{138}$, (g) Ag$_{92}$ and (h) Ag$_{68}$. Note that the screened results have been rescaled by the indicated factors and the dotted vertical lines denote the Fermi level.}
\label{fig: AbsAir}
\end{figure}

We first study the optical properties of Ag nanoparticles in air ($\epsilon_1=1$) and consider the effect of internal screening by the d-electrons. The left column of Fig.~\ref{fig: AbsAir} shows the optical absorption cross section $\sigma$ of different Ag nanoparticles calculated with and without d-electron screening (see methods section for details). For all nanoparticles, the spectrum is dominated by a single LSP peak. Inclusion of d-electron screening redshifts the LSP frequency compared to its unscreened value because screening weakens the interaction between conduction electrons facilitating the excitation of a collective oscillation. Interestingly, the size of the redshift depends sensitively on the nanoparticle radius, see Table~\ref{tab: results}. For example, a  redshift of $2.1$~eV is found for Ag$_{92}$, but for Ag$_{138}$ it is only $0.1$~eV. As the nanoparticle size increases, the plasmon energy decreases non-monotonically. 

For Ag$_{92}$, the LSP peak breaks into multiple peaks when d-electron screening is included. This Landau fragmentation is caused by the coupling of the collective LSP excitation to electron-hole pair excitations that have a similar energy~\cite{Lerme1998,Yannouleas1993}.

\begin{table}
\caption{Effect of d-electron screening on the energy $\hbar\omega_P$ of the localized surface plasmon in Ag nanoparticles in air. All energies in eV.}
\centering
\begin{tabular}{ |c|c|c|c| }
\hline
  & with d-electron screening & without d-electron screening & Redshift \\
\hline
Ag$_{254}$  & 3.4  & 4.6 & 1.2 \\
Ag$_{138}$  & 3.1 & 3.2 & 0.1 \\
Ag$_{92}$  & 3.6 & 5.5 & 2.1\\
Ag$_{68}$  & 3.8 & 5.2 & 1.4 \\
\hline 
\end{tabular}
\label{tab: results}
\end{table}

\begin{figure}[h!]
\centering
\includegraphics[width=0.35\textwidth]{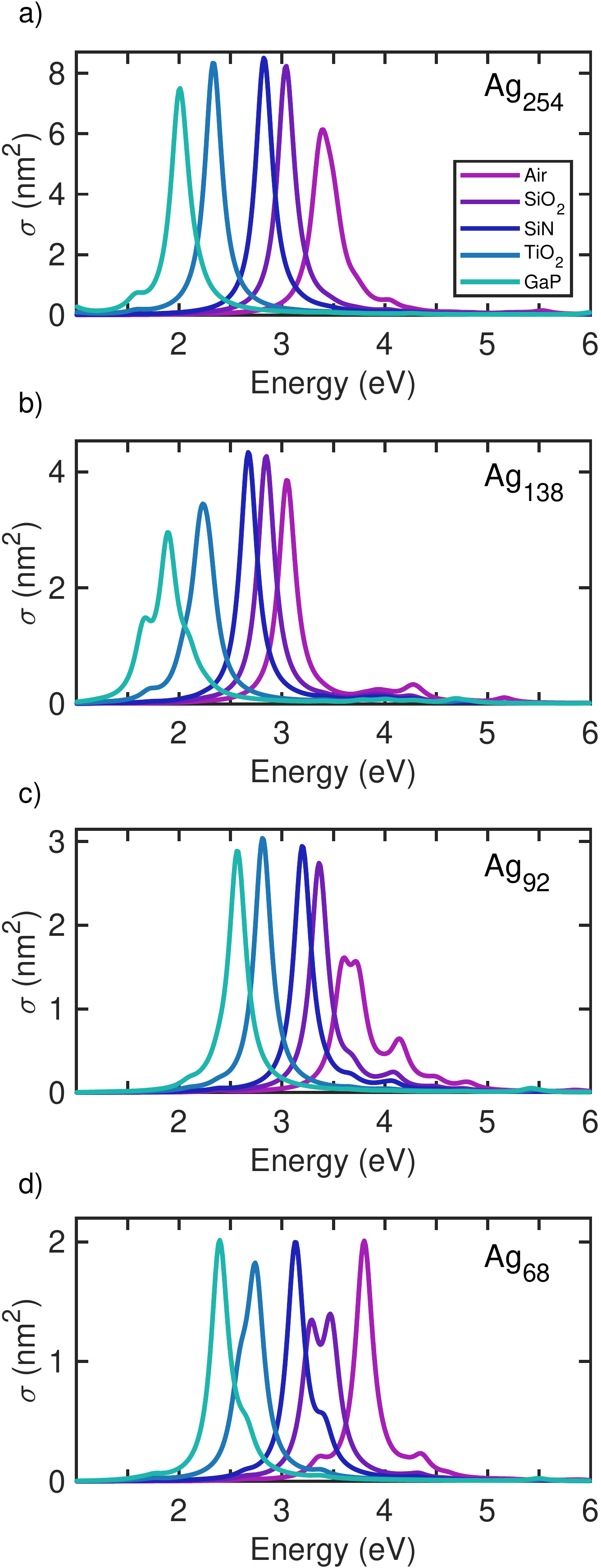}
\caption[Absorption spectra]{Absorption spectra of a) Ag$_{254}$, b) Ag$_{138}$, c) Ag$_{92}$ and d) Ag$_{68}$ embedded in SiO$_2$ ($\epsilon_1=2.16$ from Ref.\citenum{Campos2019}), SiN ($\epsilon_1=3.2$ from Ref.~\citenum{Raza2015}), TiO$_2$ ($\epsilon_1=8.2$ from Ref.~\citenum{DeVore1951}) and GaP ($\epsilon_1=20.2$ from Ref.~\citenum{Jellison1992}).}
\label{fig:AbsDiel}
\end{figure}

Next, we study the effect of embedding Ag nanoparticles in insulating host materials. In particular, we consider the following hosts: silicon dioxide ($\text{SiO}_{\rm 2}$), silicon nitride (SiN), titanium dioxide (TiO$_2$) and gallium phosphide (GaP). These materials are often used in experiments\cite{Raza2015,Mukherjee2013,Li2016,Campos2019} and cover a wide range of dielectric constants, see caption of Fig.~\ref{fig:AbsDiel}. Fig.~\ref{fig:AbsDiel} compares the optical absorption spectra of embedded Ag nanoparticles with those obtained in air. Internal screening by d-electrons is included in all calculations. 

\begin{figure}[h!]
\centering
\includegraphics[width=0.4\textwidth]{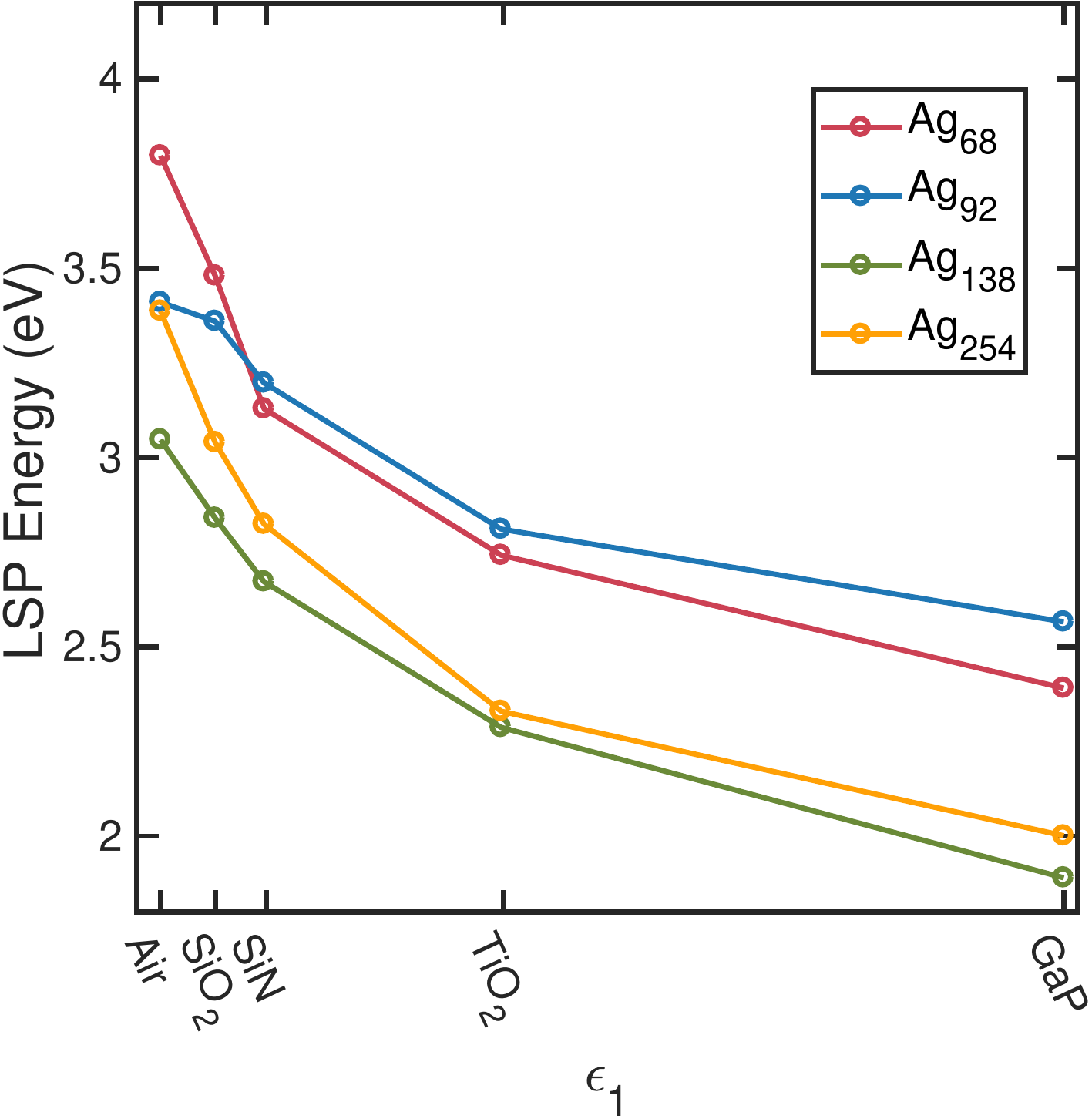}
\caption[Energy of the localized surface plasmon (LSP) versus dielectric constant of the environment]{Localized surface plasmon energy of silver nanoparticles embedded in SiO$_2$ ($\epsilon_1=2.16$ from Ref.\citenum{Campos2019}), SiN ($\epsilon_1=3.2$ from Ref.~\citenum{Raza2015}), TiO$_2$ ($\epsilon_1=8.2$ from Ref.~\citenum{DeVore1951}) and GaP ($\epsilon_1=20.2$ from Ref.~\citenum{Jellison1992})}
\label{fig: Eps}
\end{figure}

We observe that the LSP frequency of embedded nanoparticles is redshifted compared to the result in air. This is again a consequence of enhanced screening due to the presence of the dielectric environment which further weakens the effective interaction between conduction electrons in the nanoparticle. Unsurprisingly, the redshift increases with the value of the environment dielectric constant $\epsilon_1$, see Fig.~\ref{fig: Eps}. In some cases the LSP peak splits into multiple peaks upon embedding. This fragmentation occurs for Ag$_{138}$ in $\text{TiO}_{\rm 2}$ and GaP and Ag$_{68}$ in $\text{SiO}_{\rm 2}$.

Next, we compare our findings to previous theoretical and experimental studies of small Ag nanoparticles. Scholl and coworkers~\cite{Scholl2012} measured LSP energies for a range of nanoparticle sizes using electron energy loss spectroscopy. In their experiments, the Ag nanoparticles are deposited on either carbon films or SiO$_2$ substrates. For the smallest systems (with diameters of about 2~nm) they observe LSP energies that are blueshifted by about 0.5~eV from the classical LSP energy found in large nanoparticles. Specifically, LSP energies in the range of 3.6-3.8~eV are observed for the smallest nanoparticles. These LSP energies are consistent with our calculations when d-electron screening is included. Interestingly, Scholl et al. also find that the LSP energies do not approach the large nanoparticle limit in a monotonic fashion. Instead, there is significant scatter on the order of several tenths of an eV. Again, this is similar to the non-monotonic behaviour in our calculations.

To explain their findings, Scholl et al.~\cite{Scholl2012} use a theoretical model developed by Genzel, Martin and Kreibig~\cite{Genzel1975}. In this semiclassical approach, the frequency-dependent bulk dielectric function of silver is replaced by an expression that includes transitions between nanoparticle states as additional Lorentz oscillator terms. In the work of Scholl et al., these transitions are obtained from an infinite spherical well model. The resulting dielectric function is then used in a classical expression for the nanoparticle absorption spectrum. The resulting LSP energies are in good agreement with the measured ones. In particular, the blue-shift at small particle sizes and the non-monotonic behaviour are correctly reproduced. Scholl et al. interpret the non-monotonic behaviour as a consequence of transitions from occupied states with high binding energies to unoccupied states near the Fermi level. These transitions play a more dominant role in small nanoparticles: at specific radii, these transitions can strongly influence the dipolar resonance condition and therefore result in significant shifts of the LSP energy. The same theoretical approach was also recently used by Saavedra and coworkers~\cite{Saavedra2016}. The good agreement between the semi-classical model and the experimental measurement shows that a detailed description of nanoparticle transitions is needed to describe optical absorption of small Ag nanoparticles. Of course, these transitions are also captured by our fully quantum-mechanical approach. Importantly, we do not make an infinite spherical well approximation, but instead use a more accurate jellium model which allows spill-out of the electron density.

A non-monotonic behaviour of the LSP energy was also observed by L\"unskens and coworkers~\cite{Lunskens2015} who used surface second harmonic generation spectroscopy to study Ag cluster consisting of 55 atoms or less. In general, they find a blue-shift of several tenths of an eV as the cluster size is reduced. 

Yu and coworkers~\cite{Yu2018} measured optical spectra of Ag cluster with up to 120 atoms. They find that the spectra are dominated by a single LSP peak if the cluster contains more than 20 atoms. They also observe a non-monotonic behaviour of the LSP energy as a function of cluster size. Overall, the LSP energies of small clusters are blue-shifted from the large nanoparticle result by several tenths of an eV. The LSP energies that are observed for these nanoparticles lie between 3.7 eV and 3.9 eV.

Tiggesb\"aumker and coworkers~\cite{Tiggesbaumker1993} used photodepletion spectroscopy to study the LSP energy of ionic Ag clusters with up to 70 atoms. They also report a blue-shift as the cluster size is decreased with some non-monotonic behaviour. For the largest cluster, Ag$^{+}_{70}$, they find an LSP energy of 3.77~eV which is in good quantitative agreement with our result for Ag$_{68}$. Similar results were obtained by Charle and coworkers~\cite{Charle1989}.

Tunability of LSP energies by nanoparticle size and also through environmental screening was demonstrated by Jensen and coworkers~\cite{Jensen2000}. By systematically increasing the thickness of the SiO$_x$ encapsulation they could control the red-shift of the LSP energy. This is consistent with our finding that the environmental screening reduces the LSP energy. Similar results were obtained by Hilger and coworkers~\cite{Hilger2000}.

\subsection{Hot carrier properties of silver nanoparticles}

The right column of Figure~\ref{fig: AbsAir} shows the energy distribution of hot carriers that are generated per unit time from the LSP decay in Ag nanoparticles of different sizes in air (obtained by evaluating Eq.~\eqref{eq:Fermi}). Results from calculations with and without d-electron screening are compared. The hot-carrier distributions exhibit sharp peaks reflecting the discreteness of the energy level spectrum of the small nanoparticles under consideration. Energy conservation requires that peaks in the hot hole and hot electron distributions that originate from the same decay process are separated by the LSP energy (which itself depends on the nanoparticle size and inclusion of d-electron screening as discussed above). 

For applications, it is often important to know how the LSP energy is distributed among the hot electron and the hot hole. We find that this depends sensitively on the inclusion of d-electron screening: when d-electron screening is neglected, energetic holes and less energetic electrons are produced (except in Ag$_{138}$), while inclusion of d-electron screening favors the generation of hot electrons (and somewhat less energetic holes). Inclusion of d-electron screening also results in drastic changes in the magnitude of hot-carrier rates. In particular, the screened hot-carrier rates in Ag$_{68}$ and Ag$_{138}$ are more than one order of magnitude smaller than the unscreened ones, while they are one order of magnitude larger in Ag$_{92}$.

\begin{figure}[h!]
\centering
\includegraphics[width=0.85\textwidth]{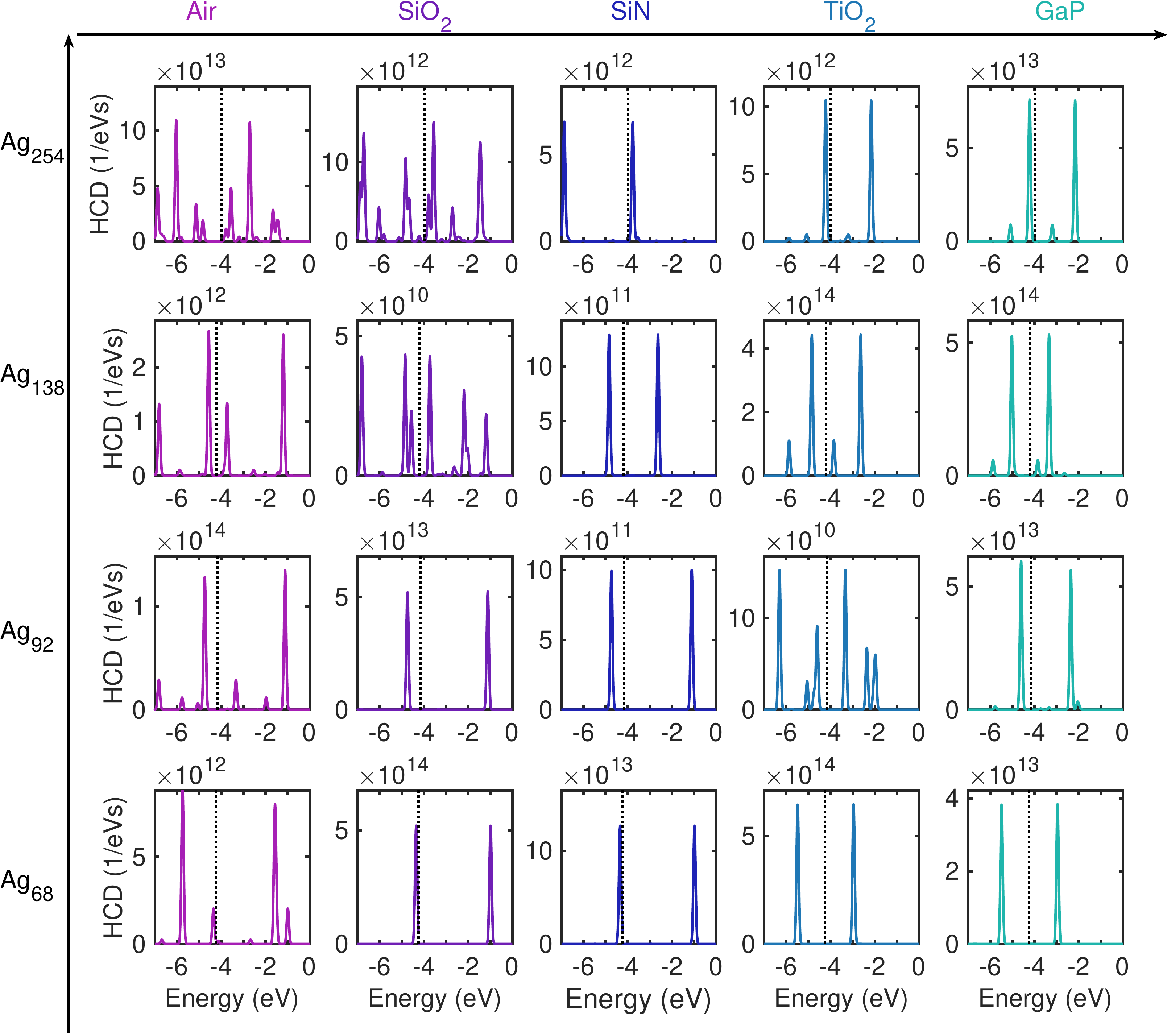}
\caption[Hot carrier distributions]{Plasmon-induced hot-carrier distributions of silver nanoparticles embedded in different host materials.}
\label{fig: HCDDiel}
\end{figure}

Figure~\ref{fig: HCDDiel} shows the hot-carrier distributions of embedded Ag nanoparticles (including internal screening by d-electrons). These distributions exhibit a large number of peaks for large nanoparticles in weakly screening environments (see, for example, Ag$_{254}$ and Ag$_{138}$ in air or $\text{SiO}_{\rm 2}$), while only a few peaks are found for nanoparticles in GaP. In most systems, the generation of energetic electrons is favored compared to energetic holes and the energy of hot carriers (measured with respect to the Fermi level) is generally larger in weakly screening environments.

We have also calculated the hot carrier distributions using a semiclassical approach (see Supplementary Information). The semiclassical hot carrier rates have a similar shape as the fully quantum-mechanical ones, but a significantly larger magnitude. This is caused by (i) the actual transition dipole moments of LSPs in small nanoparticles being smaller than the classical result and (ii) the spectral weight of the LSP being reduced due to coupling to electron-hole pair excitations (see detailed discussion in Ref.~\citenum{RomanCastellanos2019}).

\begin{figure}[h!]
\centering
\includegraphics[width=0.5\textwidth]{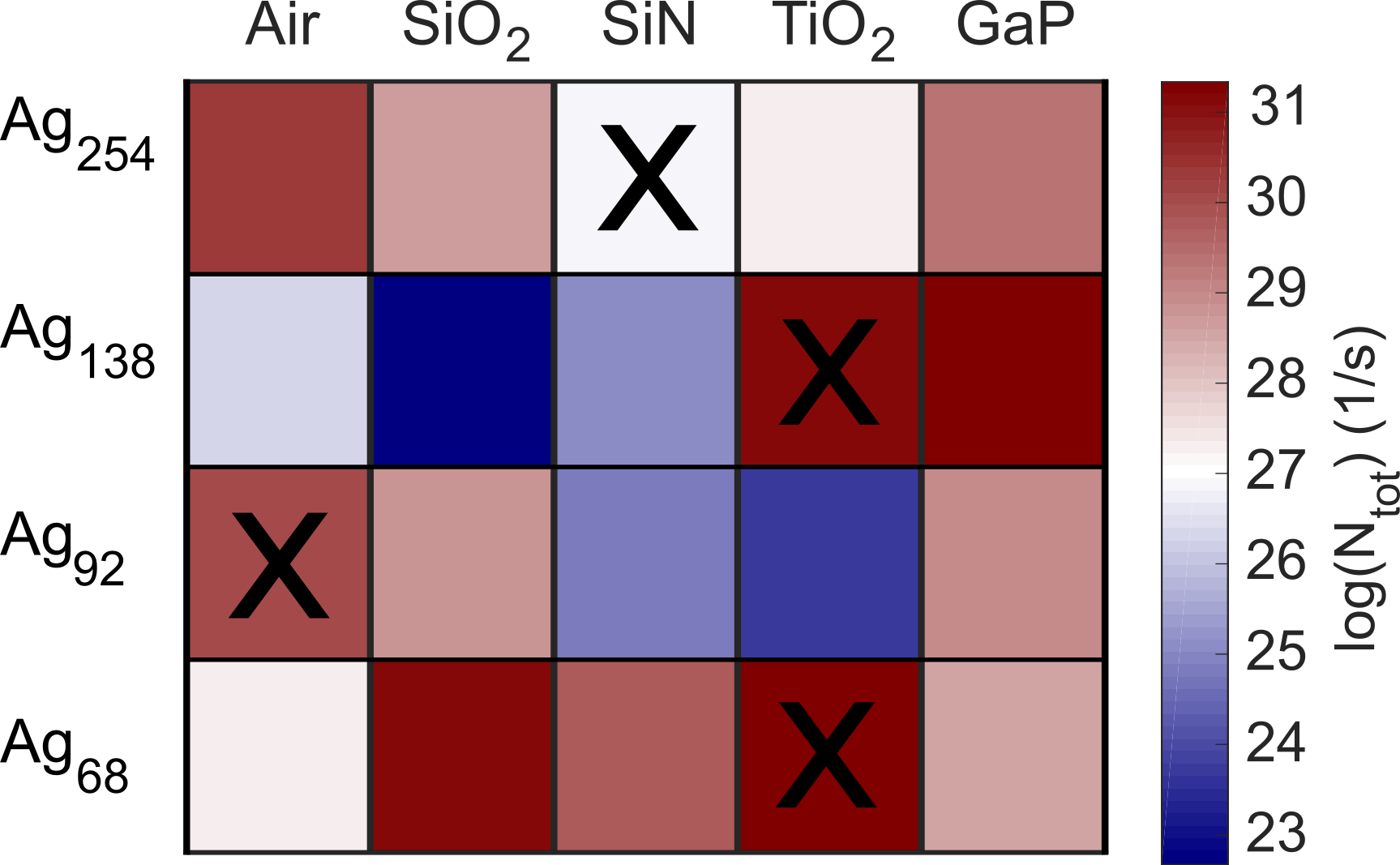}
\caption[Matrix of the total number of hot carriers in different dielectric environments]{Total hot-carrier generation rates of embedded Ag nanoparticles. The crosses indicate the environment where the highest hot-carrier rates are expected based on the analysis of the nanoparticle joint density of states, see Figs.~\ref{fig: JDOS&g2} a) and b).}
\label{fig: MatrixTotalLogHCD}
\end{figure}

Figure~\ref{fig: MatrixTotalLogHCD} shows the total number of hot carriers produced per unit time in embedded Ag nanoparticles. Large hot-carrier rates are generally obtained in environments with large dielectric constants (in particular, GaP and TiO$_2$), but for Ag$_{92}$ and Ag$_{254}$ the maximum rate is actually obtained in air.

\begin{figure}[h!]
\centering
\includegraphics[width=0.6\textwidth]{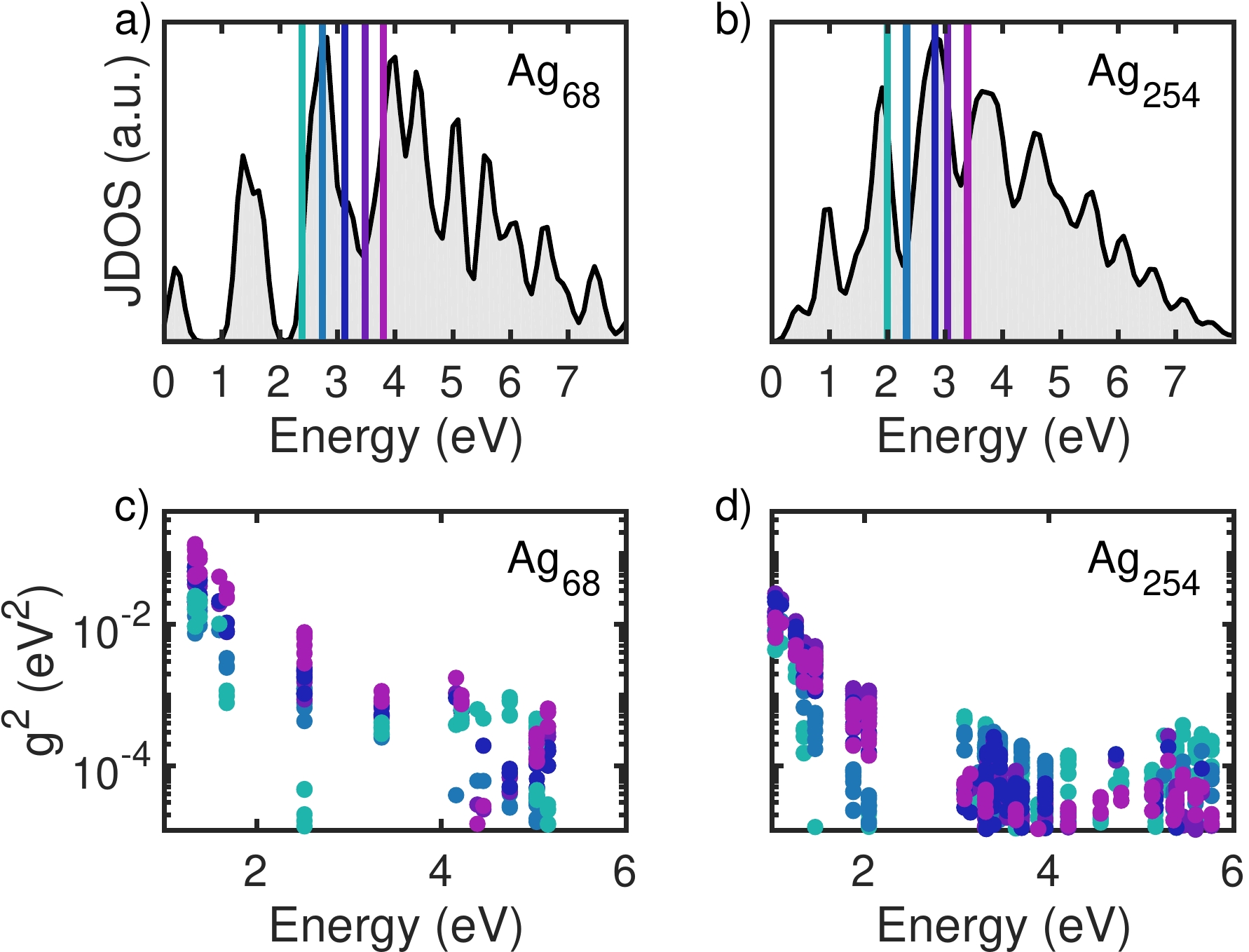}
\caption[Joint density of bound states and electron-plasmon coupling versus the single particle energy difference.]{Top: Joint density of bound states of a) Ag$_{68}$ and b) Ag$_{254}$. The vertical lines denote the localized surface plasmon energy of the nanoparticles embedded in different host materials (from left to right: air, SiN, SiO$_{2}$, TiO$_2$ and GaP). Bottom: Electron-plasmon coupling in c) Ag$_{68}$ and d) Ag$_{254}$ in different host materials as a function of transition energy.}
\label{fig: JDOS&g2}
\end{figure}

To understand the observed trends, we separately analyze the two ingredients that determine hot-carrier rates according to Fermi's golden rule, see Eq.~\eqref{eq:Fermi}: the number of available energy-conserving transitions which is described by the joint density of states and the electron-plasmon coupling. For most hot-carrier applications, only hot electrons in bound states are relevant and we therefore restrict ourselves to the joint density of bound states (JDOBS). Figs.~\ref{fig: JDOS&g2} a) and b) show the JDOBS of Ag$_{68}$ and Ag$_{254}$ as function of the excitation energy. Because of the discreteness of the electronic energy levels, the JDOBS curves are oscillatory, but they exhibit an overall maximum near 3~eV. Therefore, the number of energy-conserving transitions is maximized when the LSP energy (denoted by vertical lines in the figure) coincides with the maximum of the JDOBS. For Ag$_{68}$, this is achieved when the nanoparticle is embedded in TiO$_2$, while for Ag$_{254}$ embedding in SiN is required. The environments that maximize the number of available transitions are denoted by crosses in Fig.~\ref{fig: MatrixTotalLogHCD} and we observe that these environments indeed yield very large hot-carrier rates. The oscillatory nature of the JDOBS also explains why small changes in the environmental screening (and therefore in the LSP energy) can lead to large changes in hot-carrier rates.

Considering next the electron-plasmon coupling, one would naively expect that increased screening would reduce $g_{vc}$ as the effective electron-electron interaction in Eq.~\eqref{eq:coupling} is weakened. Figs.~\ref{fig: JDOS&g2} c) and d) show that  this trend is indeed followed in Ag$_{68}$ and Ag$_{254}$. However, this reduction in $g_{vc}$ is not directly relevant to hot-carrier rates because the concomitant reduction of the LSP frequency leads to the excitation of different $vc$-transitions with smaller energies. As the excitation energy decreases, the electron-plasmon coupling increases, see Figs.~\ref{fig: JDOS&g2} c) and d), and this explains the observed large hot-carrier rates in GaP in Fig.~\ref{fig: MatrixTotalLogHCD}.

\section{Conclusions}

We have developed a quantum-mechanical approach for describing hot carriers resulting from the decay of localized surface plasmons in small silver nanoparticles that are embedded in dielectric media. Dielectric screening by the nanoparticle environment and by Ag d-electrons is taken into account by means of an effective electron-electron interaction which is used to calculate electron-electron interaction matrix elements and electron-plasmon couplings. We find that hot-carrier generation rates depend sensitively on the environmental and internal screening. We demonstrate that hot-carrier production can be maximized by choosing the host material such that the LSP energy of the embedded nanoparticle coincides with the maximum of its joint density of states. Moreover, high hot-carrier generation rates are achieved in host materials with very large dielectric constants as the concomitant small LSP energies result in large electron-plasmon couplings. In this case, however, the hot carriers are less energetic. These insights can be used as design rules for creating efficient hot-carrier devices and open up the possibility of tailoring hot-carrier properties by dielectric engineering.

\section{Competing interests}
The authors declare no competing interests.

\section{Supporting information}
Dependence of the absorption spectrum on $\epsilon_{d}$.\\
Semiclassical hot carrier distributions.

\newpage

\begin{acknowledgement}
The authors acknowledge support from the Thomas Young Centre under grant no. TYC-101. This work was supported through a studentship in the Centre for Doctoral Training on Theory and Simulation of Materials at Imperial College London funded by the EPSRC (EP/L015579/1) and  through EPSRC projects EP/L024926/1 and EP/L027151/1. Support by the Science Foundation Ireland (SFI) under grant 18/RP/6236 is gratefully acknowledged.
\end{acknowledgement}

\begin{figure}
\caption*{TOC Graphic}
\includegraphics[scale=0.10]{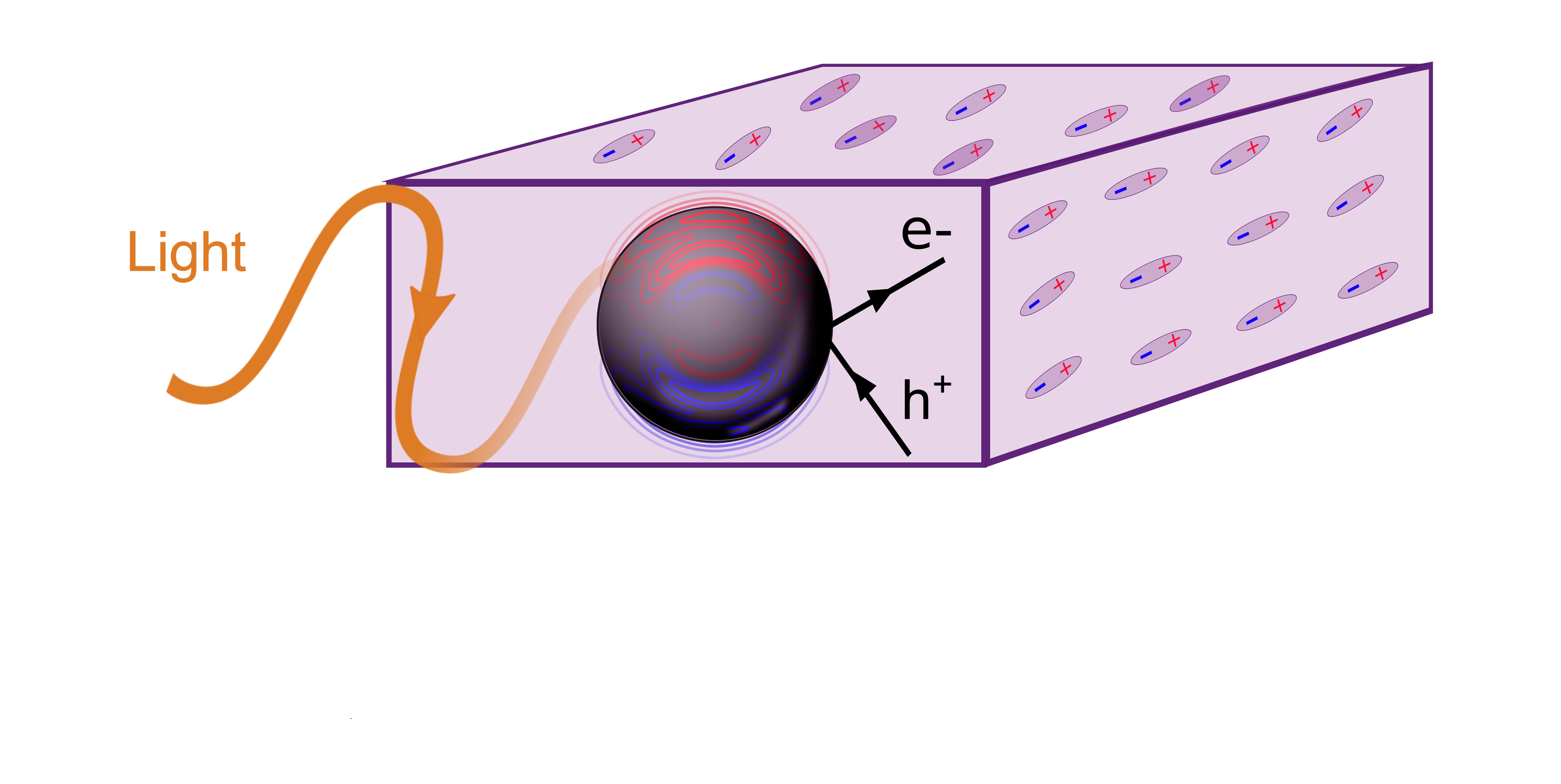}
\end{figure}

%\section{Supporting Information}

%\bibliography{libraryCLEAN.bib}

\bibliography{main.bbl}

\end{document}